\documentclass[journal]{IEEEtran}
\ifCLASSINFOpdf
\else
\fi
\usepackage{sidecap}
\usepackage{caption}
\usepackage{graphicx}
\usepackage{subfig}
\usepackage{tikz}
\usepackage{multirow}
\usepackage{amsmath}
\usepackage{booktabs}
\usepackage{textcomp}

\begin{document}
\title{Reconfigurable photonic integrated mode (de)multiplexer for SDM fiber transmission}

\author{Daniele Melati, Andrea Alippi, and Andrea Melloni\\% <-this % stops a space
\small Dipartimento di Elettronica, Informazione e Bioingegneria, Politecnico di Milano, 20133 Milano, Italy\\%
daniele.melati@polimi.it.}% <-this % stops a space
\maketitle

\begin{abstract}
A photonic integrated circuit for mode multiplexing and demultiplexing in a few-mode fiber is presented and demonstrated. Two 10 Gbit/s channels at the same wavelength and polarization are simultaneously transmitted over modes LP$_{01}$ and LP$_{11a}$ of a few-mode fiber exploiting the integrated mode MUX and DEMUX. The proposed Indium-Phosphide-based circuits have a good coupling efficiency with fiber modes with mode-dependant loss smaller than 1 dB. Measured mode excitation cross-talk is as low as -20 dB and a channel cross-talk after propagation and demultiplexing of -15 dB is achieved. An operational bandwidth of the full transmission system of at least 10 nm is demonstrated. Both mode MUX and DEMUX are fully reconfigurable and allow a dynamic switch of channel routing in the transmission system. 
\end{abstract}


% \fi
\IEEEpeerreviewmaketitle

\section{Introduction}

\IEEEPARstart{C}{onstant} progress of optical technologies allowed to increase the data-transport capacity of each single optical fiber through the exploitation of multiplexing in phase, time, wavelength and polarization. Nonetheless, fiber capacity limit is expected to be around the corner \cite{richardson2013space, ellis2010approaching}. Research has hence focused on the exploration of 'space' as the fifth physical dimension for signals multiplexing which could avoid capacity crunch \cite{winzer2014making}. Spatial, or mode, multiplexing allows to conveniently increase the transmission capacity of optical transport systems, being largely more energy-efficient if compared to single-fiber systems at high capacity \cite{winzer2011energy}.

%apertura SDM
Although few-mode optical fibers (FMFs) were proposed more than thirty years ago, they have recently drawn the attention as a viable solution for the implementation of Spatial Division Multiplexing (SDM) \cite{richardson2013space}, among other ideas such as multi-core fibers or single-mode fiber bundles. Compared to single-mode fibers (SMFs), FMFs increase the fiber spectral efficiency exploiting orthogonal propagating transverse modes to simultaneously convey multiple data streams, which are recovered through a receiver equipped with mode diversity capabilities  \cite{chen2014compact}. Mode multiplexing over FMFs allowed to reach transmission rates of several Terabit per second over distances up to hundreds of kilometres \cite{chen2014compact, van2014ultra}. Few-mode compatible network components such as (de)multiplexers, erbium-doped fiber amplifiers \cite{sleiffer201273} and optical add/drop multiplexers \cite{chen2012reception} have been proposed as well.

%integrazione
Many of the reported systems rely on free-space optical elements (e.g. lenses, beam splitters and phase plates) \cite{ryf2012mode,van2014ultra} and/or photonic lanterns \cite{fontaine201530} to prepare optical modes, (de)multiplex and (de)couple them with FMF modes. Although effective, these solutions can be hardly scaled-up and included in densely-packed commercial modules. Photonics integration is hence a mandatory technology for future few-mode compatible optical networks, enabling low-loss mode couplers \cite{richardson2013space, koonen2012silicon}, reducing costs, further improve power efficiency \cite{winzer2014making} and making mode-based SDM solutions competitive against single-mode networks. Alongside with pure on-chip design for data interconnect \cite{luo2014wdm, dai2013silicon}, photonic integrated circuits (PICs) have been proposed also as mode (de)multiplexers for SDM transmission applications. Typically, PICs rely on both vertical spot couplers \cite{ding2013silicon, chen2014compact} and butt coupled mode converters \cite{dai2015mode}.

%innovation
Even if the use of PICs either as mode multiplexer \cite{chen2013demonstration} or demultiplexer \cite{doerr2011proposed, fontaine2012space} for SDM transmission has already been proposed, up to our knowledge a clear demonstration of a transmission system relying exclusively on PICs for the implementation of both functions has not yet been reported. In this work, a PIC providing both two-mode multiplexing at the transmitter and demultiplexing at the receiver is proposed and demonstrated. Circuit performances are evaluated with both simulations and experimental characterizations at 10 Gbit/s. The integrated transmitter is used to selectively excite modes LP$_{01}$ and LP$_{11a}$ of a FMF which are demultiplexed by the receiver after propagation. A few-meter-long fiber is used in this work in order to demonstrate mode (de)multipexing functionality of the proposed PIC. A coupling efficiency of about -6 dB is achieved with mode-dependant loss smaller than 1 dB. On-chip mode cross-talk as low as - 20 dB is measured while a channel cross-talk of about -15 dB is achieved after fiber propagation and demultiplexing. Moreover, reconfiguration capabilities of the mode (de)multiplexer allow to dynamically change signal routing at both transmitter and receiver. An operational bandwidth of at least 10 nm is demonstrated as well, making the design compatible with flexible networks and WDM transmission. Being the proposed PICs actually polarization insensitive, any polarization control element is avoided in the demonstrated transmission system. PICs are realized in Indium-Phosphide-based technology integrating PIN photodiodes at the receiver and are suitable for laser sources and SOAs integration for a fully integrated SDM transceiver. Lastly, two-mode SDM 10Gbit/s transmission over few meters of a FMF is demonstrated exploiting the proposed PIC as mode multiplexer and demultiplexer.

%content
This paper is organized as follows. Section \ref{sec: concept} presents the design of the circuit, the reconfiguration principle based on amplitude and phase controllers and the coupling efficiency issues with fiber modes LP$_{01}$ and LP$_{11a}$. Section \ref{sec: realization} reports the experimental characterization of both PIC operation and SDM transmission system exploiting the integrated mode (DE)MUX and a few-mode fiber. Circuit reconfiguration and bandwidth measurement are reported as well. In Section \ref{sec: SDM} the integrated mode MUX and DEMUX are exploited for the transmission of two 10 Gbit/s channels over a few-mode fiber.

\section{Concept and circuit design}
\label{sec: concept}
\begin{center}
\begin{figure}[ht]
\centering
\includegraphics[keepaspectratio=true, width=0.9\columnwidth]{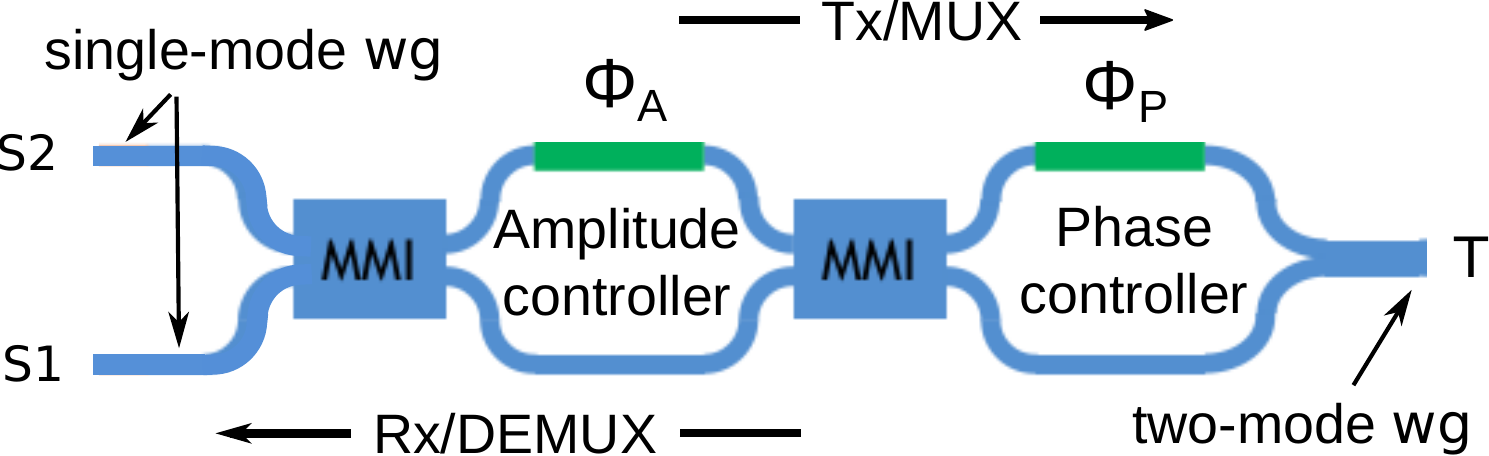}
\caption{Schematic of the integrated mode (de)multiplexer. Two single mode waveguides are coupled to the fundamental and first higher order mode of a two-mode waveguide through a tuneable, linear, and fully reconfigurable circuit.}
\label{fig:CircuitScheme}
\end{figure}
\end{center}
%circuit description
The schematic of the proposed mode DE(MUX) is shown in figure \ref{fig:CircuitScheme}. When used as multiplexer, the circuit provides two single-mode input waveguides (S1 and S2, to be coupled to single-mode fibers) and one two-mode output waveguide (T, to be coupled to a few-mode fiber). The circuit implements a reconfigurable linear network \cite{miller2013reconfigurable} that convert the fundamental modes of the two separated single-mode waveguides into the fundamental and first higher order mode of a two-mode waveguide. In the most general form it can be implemented by cascading a tunable amplitude controller and a tunable phase controller. The amplitude controller allows to adjust the splitting ratio of the input signals in the two branches of the relative phase controller. In this work, the amplitude controller is realized by means of a tunable coupler based on a balanced Mach-Zehnder interferometer with two 2x2 MMIs and a phase shifter $\phi_{A}$. A second phase shifter is used to control the relative phase between the two branches after the tuneable coupler ($\phi_{P}$) and to excite the desired mode. A symmetric Y-branch \cite{burns1988waveguide} with a two-mode common waveguide finally combines the two branches.

%circuit operation
Since the circuit is completely reconfigurable, the two single-mode inputs can be mapped into any combination of the two modes of the output waveguide. We consider here only the simplest case where one input is coupled into either the fundamental or the first-higher-order output mode. The other input is consequently mapped on the other mode. This can be achieved by operating the amplitude controller as a 3-dB coupler ($\phi_{A} = \pi/2$). If $\phi_{P} = 0$ then the input S2 is mapped on the fundamental mode while S1 on the first order mode; if $\phi_{P} = \pi$ the opposite mapping is obtained. Fundamental and first order mode are hence used to excite modes LP$_{01}$ and LP$_{11a}$ of a FMF.  Any other combination of $\phi_{A}$ and $\phi_{P}$ allows to generate at the output any field distribution as a linear combination of the two guided modes. Demultiplexing operation is obtained by reversing inputs and outputs, that is coupling light in the two-mode waveguide T through the FMF and collecting the output signals at the two single mode waveguides S1 and S2 with two SMFs or photodiodes.
%In this case DEMUX can map any input field distribution (any linear combination of the two modes of the 4-\textmi{}m waveguide) on either IN1 or IN2, by effectively implementing also a one-tap SDM equalizer.

%multi mode section simulation
\begin{center}
\begin{figure}[t]
\centering
\includegraphics[scale=0.43]{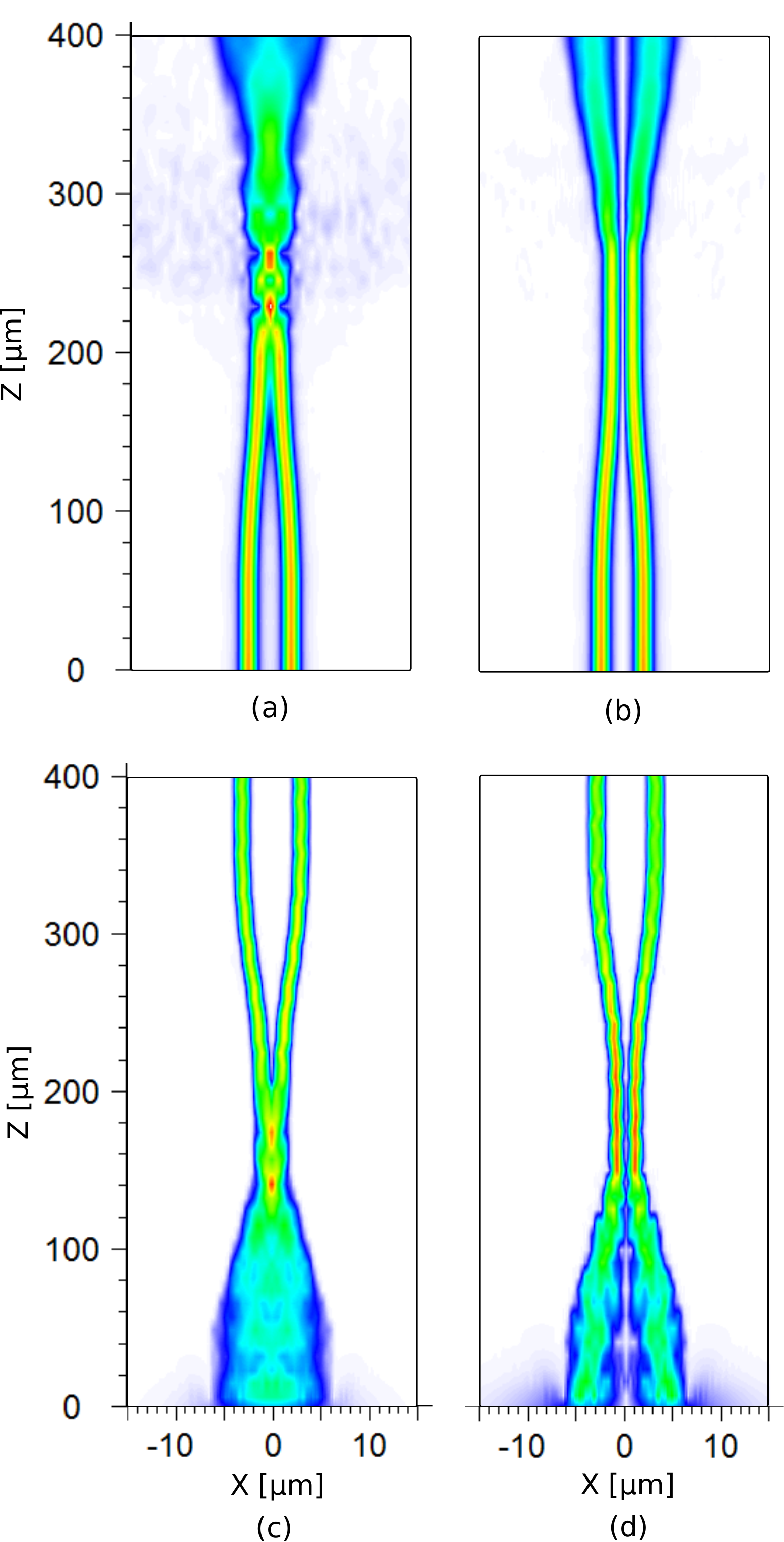}
\caption{Propagation simulations of the integrated mode (DE)MUX based on Beam Propagation Method (color scale is normalized). Z-axis refers the direction of propagation while X-axis refers the cross-section view. (a) and (b) show the propagation in the integrated mode MUX with fundamental mode inputs with 0 and $\pi$ phase difference, respectively. (c) and (d) refer to the DEMUX case, considering as input LP$_{01}$ and LP$_{11a}$ mode, respectively.}
\label{fig:PropSimulation}
\end{figure}
\end{center}

Beam Propagation Method was used to verify the circuit behaviour and asses coupling efficiency between the output waveguide and a FMF. The considered waveguide structure is rib-shaped with 1 \textmu{}m-thick InGaAsP core on top of the InP substrate, etch depth of 1.7 \textmu{}m and no top cladding \cite{smit2014introduction}. Single-mode waveguides have a width of 2\textmu{}m while output waveguide is 4-\textmu{}m-wide and allows the propagation of two guided modes. In order to provide low coupling losses, an adiabatic taper widen the waveguide width to 13 \textmu{}m before chip output facet. After the taper, fundamental output mode excites LP$_{01}$ mode of the FMF while first-order mode couples on mode LP$_{11a}$.
%Also the single mode waveguides S1 and S2 have a suitable mode adapter \cite{soares2013semi}.

The final part of the phase shifter and the two mode taper was simulated at an operational wavelength of $\lambda=1550$ nm. Figures \ref{fig:PropSimulation}(a,b) refer to the MUX operation.  At the input, the fundamental mode is excited in the two single-mode waveguides with the same amplitude since the tuneable coupler is considered as set at 3-dB state (bottom of the figures). In figure \ref{fig:PropSimulation}(a) the two inputs have the same phase and the fundamental mode is excited at the end of the taper. Simulated coupling efficiency with LP$_{01}$ mode of the FMF is about -6 dB. As mentioned before, this case corresponds to use input S2 with $\phi_{P} = 0$ or S1 with $\phi_{P} = \pi$. In figure \ref{fig:PropSimulation}(b) the two inputs have a phase difference of $\pi$ rad (input S1 with $\phi_{P} = 0$ or S2 with $\phi_{P} = \pi$) and the first higher order mode at the output is excited. Coupling efficiency with LP$_{11a}$ mode of the FMF is about -7 dB, confirming small mode-dependent losses. In both cases the insertion loss and mode conversion in the adiabatic taper are negligible. In figures \ref{fig:PropSimulation}(c,d) the DEMUX operation is considered. Input is provided by either mode LP$_{01}$, figure \ref{fig:PropSimulation}(c), or LP$_{11a}$, figure \ref{fig:PropSimulation}(d). In both cases, at the two singe-mode output the amplitude is the same in both cases while phase difference is 0 and $\pi$ rad, respectively. Simulated coupling efficiencies are the same as in the MUX operation.

%\cite{morichetti2015all}.
%------------------------------------------------

\section{Realization and experimental results}
\label{sec: realization}

\begin{center}
\begin{figure}[ht]
\centering
\includegraphics[keepaspectratio=true, width=1\columnwidth]{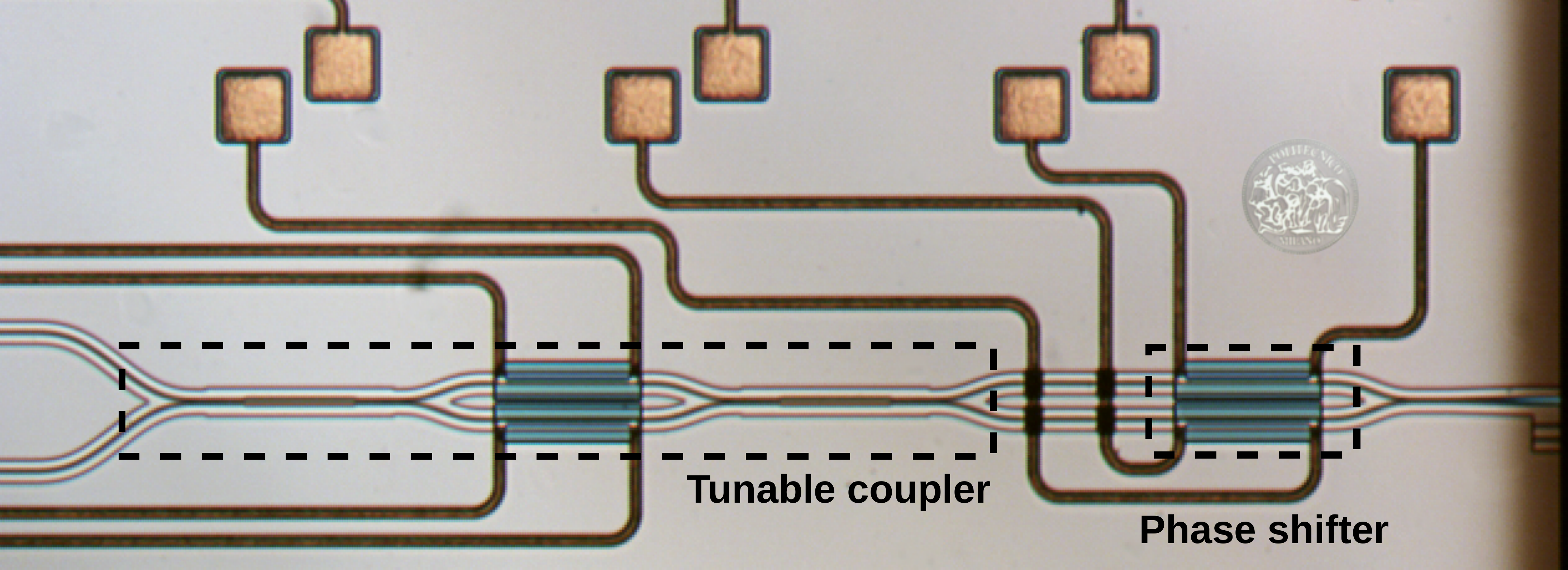}
\caption{Photograph of the realized integrated mode (de)multiplexer.}
\label{fig:photo}
\end{figure}
\end{center}

As mentioned in the previous section, the device was designed and fabricated on an InP-based generic technological platform through a Multi-Project Wafer run \cite{smit2014introduction,jeppix}. A photograph of the realized device is shown in figure \ref{fig:photo} where the tunable coupler and the phase shifter are highlighted. Input waveguides have a pitch of 250 \textmu{}m to allow coupling with a commercial 2-fiber array. Spot-size converters (not visible in the photograph) ensure an efficient light coupling between standard SMFs and the single-mode waveguides (width 2 \textmu{}m), with insertion loss lower than 2 dB and PDL smaller than 0.5 dB \cite{smit2014introduction,soares2013semi}. The 13-\textmu{}m taper can be clearly seen at chip edge on the right. Input and output facets are covered with anti-reflective coating to avoid spurious reflections. Phase shifters are realized with 200-\textmu{}m-long thermo-optic heaters with a V$_\pi$ of about 4.8 V. Integrated PIN photodiodes for direct electrical access to demultiplexed signals have been added to the devices used as DEMUX.

\begin{center}
\begin{figure}[ht]
\centering
\includegraphics[keepaspectratio=true, width=0.9\columnwidth]{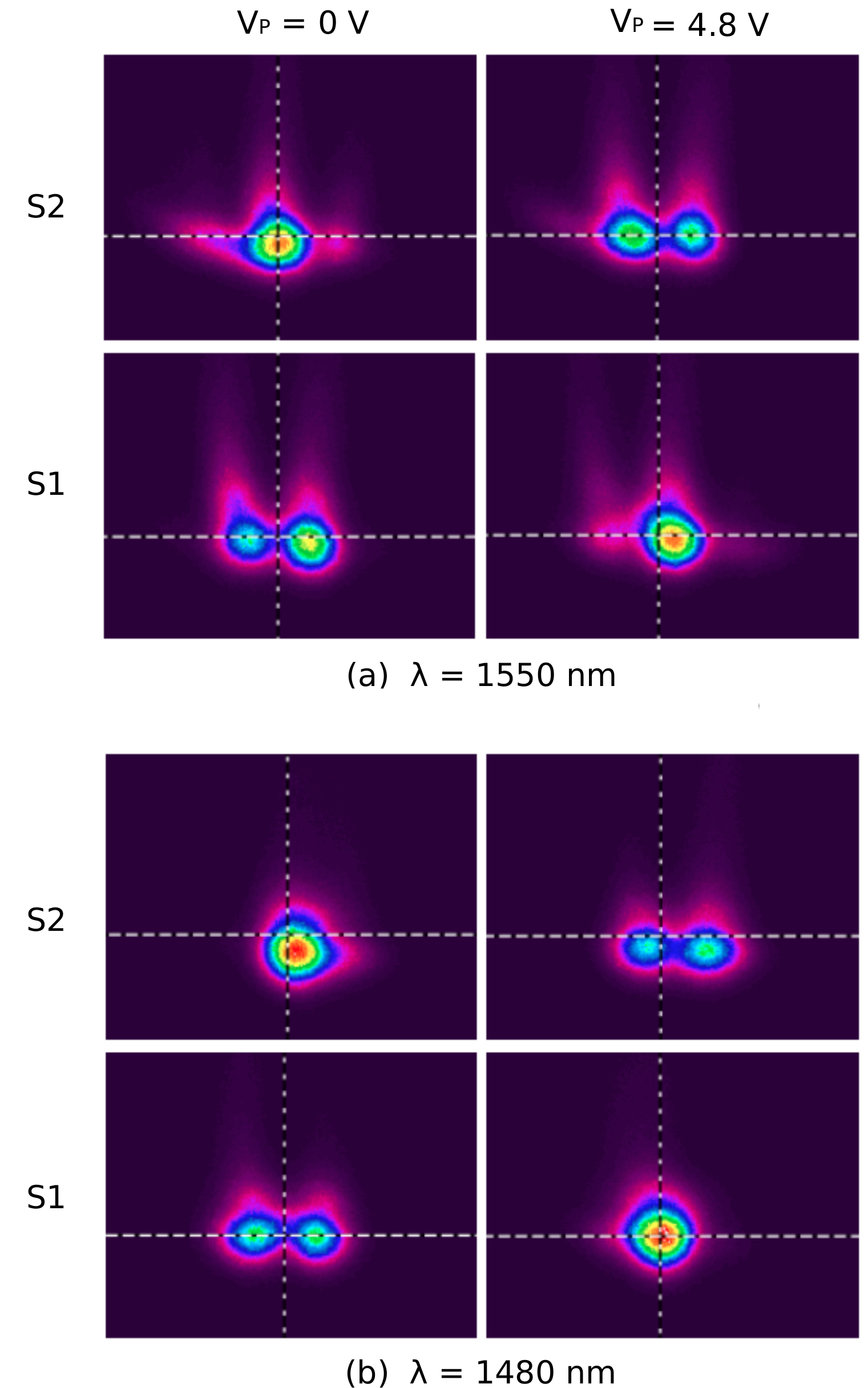}
\caption{Mode field distribution at the output of the MUX circuit for different inputs and different voltages applied at the phase controller. V$_{P}$ = 4.8 V allows to obtain a phase shift of $\phi_{P} = \pi$. Tunable coupler is set at 3-dB state with V$_{A}$ = 3.1 V ($\phi_{A} = \pi/2$). Wavelength of the input light is either (a) 1550 nm or (b) 1480 nm.}
\label{fig:mode maps}
\end{figure}
\end{center}

As a first analysis, we experimentally characterised the functionality of the MUX operation. A single-mode fiber was used to couple light through either S1 or S2 and the output field distribution was measured with an objective aligned with a CCD camera. No polarization controller was used in the setup. Results are shown in figure \ref{fig:mode maps}(a) for a wavelength $\lambda$ = 1550 nm. The tunable coupler was set at 3-dB state ($\phi_{A} = \pi/2$) applying a voltage V$_{A}$ = 3.1V. As can be clearly seen, when $\phi_{P}$ = 0 (V$_{P}$ = 0) and light is coupled to input port S2, the fundamental mode is excited at the output T while input S1 is mapped on the first-order mode. Mode cross-talk in a SMF was measured as low as -20 dB. With a $\pi$ shift of the phase controller ($\phi_{P}$ = $\pi$, V$_{P}$ = 4.8 V) the configuration is switched: S2 is mapped on the first-order output mode and S1 on the fundamental mode. Since the circuit is completely balanced the operational bandwidth of the MUX is extremely wide. This is demonstrated in figure \ref{fig:mode maps}(b) where wavelength of the input light was changed to 1480 nm while keeping circuit settings unchanged. As in the previous case, V$_{P}$ = 0 maps input S2 on the fundamental mode and S1 on the first-order mode while V$_{P}$ = 4.8 V switches the MUX configuration.

\begin{center}
\begin{figure}[ht]
\centering
\includegraphics[scale=0.46 ]{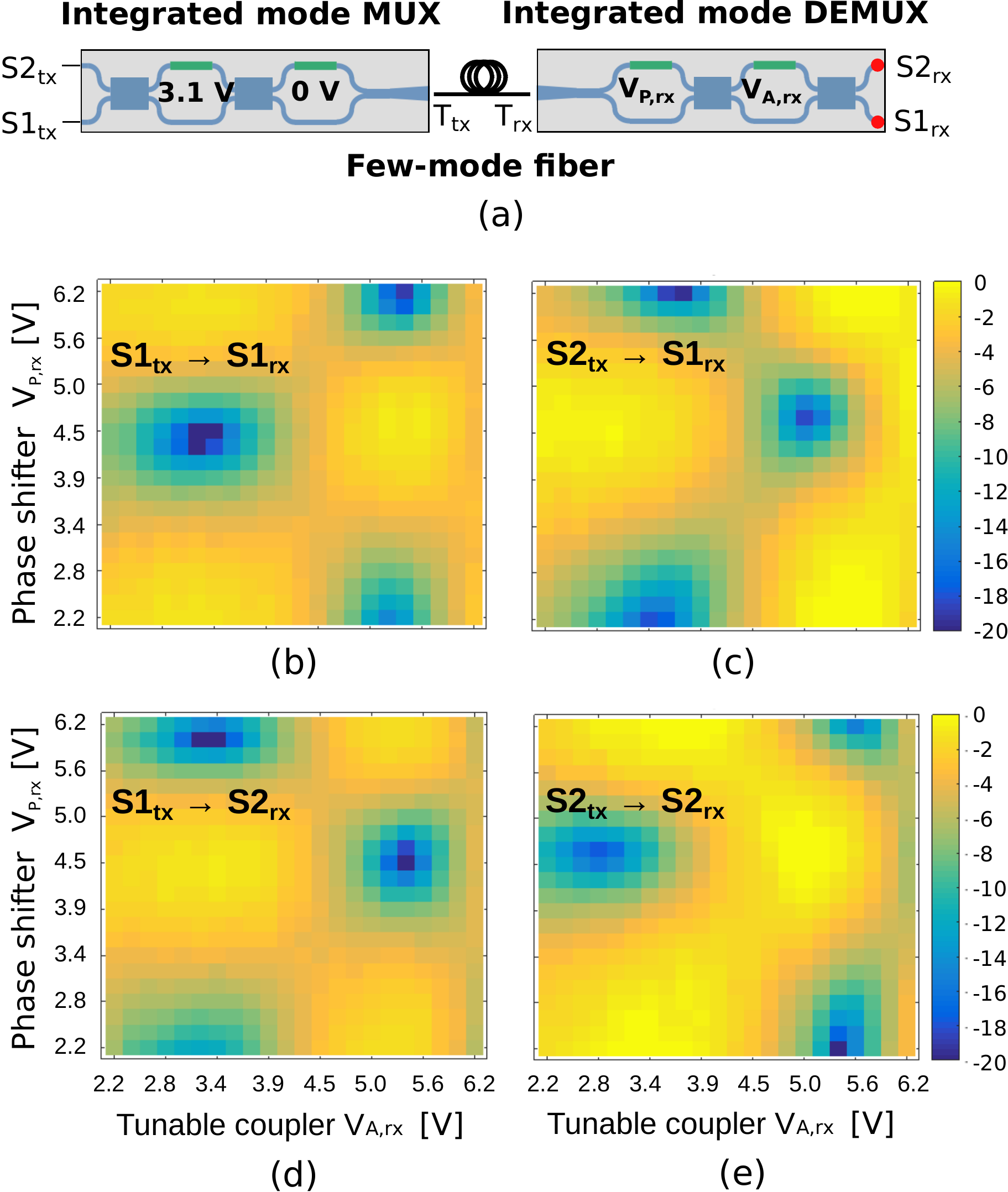}
\caption{(a) Sketch of the SDM transmission test setup with integrated mode MUX and DEMUX. The demultiplexer output waveguides S1$_{\mathrm{rx}}$ and S2$_{\mathrm{rx}}$ are equipped with integrated PIN photodiodes. (b-e) Normalized transmission from S1$_{\mathrm{tx}}$ and S2$_{\mathrm{tx}}$ to S1$_{\mathrm{rx}}$ and S2$_{\mathrm{rx}}$ as function of the voltage applied to DEMUX thermo-optic heaters (V$_{\mathrm{A,rx}}$ and V$_{\mathrm{P,rx}}$). The MUX is set at V$_{\mathrm{A,tx}}$ = 3.1 V,  V$_{\mathrm{P,tx}}$ = 0 V for all the reported results.}
\label{fig:switch maps}
\end{figure}
\end{center}
\begin{table}[ht]
\caption{Normalized transmission from S1$_{\mathrm{tx}}$ and  S2$_{\mathrm{rx}}$ to S1$_{\mathrm{rx}}$ and S2$_{\mathrm{rx}}$ for two different DEMUX configurations. Channels cross-talk at the two output photodiodes is reported as well.}
\begin{center}

\begin{tabular}{ccp{2cm}p{2cm}}
\toprule
 & & V$_{\mathrm{A,rx}}$ = 3.0V V$_{\mathrm{P,rx}}$ = 4.5V & V$_{\mathrm{A,rx}}$ = 5.2V V$_{\mathrm{P,rx}}$ = 4.5V \\
\midrule
\multirow{2}*{S1$_{\mathrm{tx}}$} & S1$_{\mathrm{rx}}$ & -16.3 dB & 0 dB \\
		   &  S2$_{\mathrm{rx}}$ & 0 dB & -16.3 dB \\ \hline
\\
\multirow{2}*{S2$_{\mathrm{tx}}$} &  S1$_{\mathrm{rx}}$ & -0.9 dB & -15.2 dB \\ 
		   &  S2$_{\mathrm{rx}}$ & -14.4 dB & -0.7 dB \\  \hline
\\
\multirow{2}*{X-talk} &  S1$_{\mathrm{rx}}$ & -15.4 dB & -15.2 dB \\
		      &  S2$_{\mathrm{rx}}$ & -14.5 dB & -15.6 dB \\ 
\bottomrule
\end{tabular}
\end{center}
\label{tab:table 1}
\end{table}

Integrated MUX and DEMUX circuits were hence exploited to realize a full SDM transmission system. The sketch of the experimental setup is shown in figure \ref{fig:switch maps}(a). A 2-meter-long graded-index FMF with mode field diameter of 11\textmu{}m was coupled to the output T$_{\mathrm{tx}}$ of the MUX and the input T$_{\mathrm{rx}}$ of the DEMUX. Angular alignment of the fiber to DEMUX input was carefully controlled in order to ensure good coupling between mode LP$_{11a}$ and first-order input mode of the DEMUX. As in the previous case, no polarization controller was used in the setup. Signal at wavelength of 1540 nm was generated and coupled to either MUX's input S1$_{\mathrm{tx}}$ or S2$_{\mathrm{tx}}$ through a standard SMF. The MUX configuration was fixed at V$_{\mathrm{A,tx}}$ = 3.1 V and V$_{\mathrm{P,tx}}$ = 0, mapping input S1$_{\mathrm{tx}}$ on the first-order mode and S2$_{\mathrm{tx}}$ on the fundamental mode at the MUX's output T$_{\mathrm{tx}}$ (see figures \ref{fig:mode maps}(a) and (b)). Acting on the DEMUX it is possible to recover the two input signals as well as perform system reconfiguration. Output signals were measured at the DEMUX's outputs S1$_{\mathrm{rx}}$ and S2$_{\mathrm{rx}}$ terminated with integrated PIN photodiodes. All input/output port combinations were investigated while changing the DEMUX setting through the integrated thermo-optic heaters. In particular, the voltage for both the tuneable coupler (V$_{\mathrm{A,rx}}$) and phase controller (V$_{\mathrm{P,rx}}$) of the DEMUX were swept between 2.2 V and 6.2 V while measuring input/output system transmission.

\begin{center}
\begin{figure}[b]
\centering
\includegraphics[keepaspectratio=true, width=0.9\columnwidth]{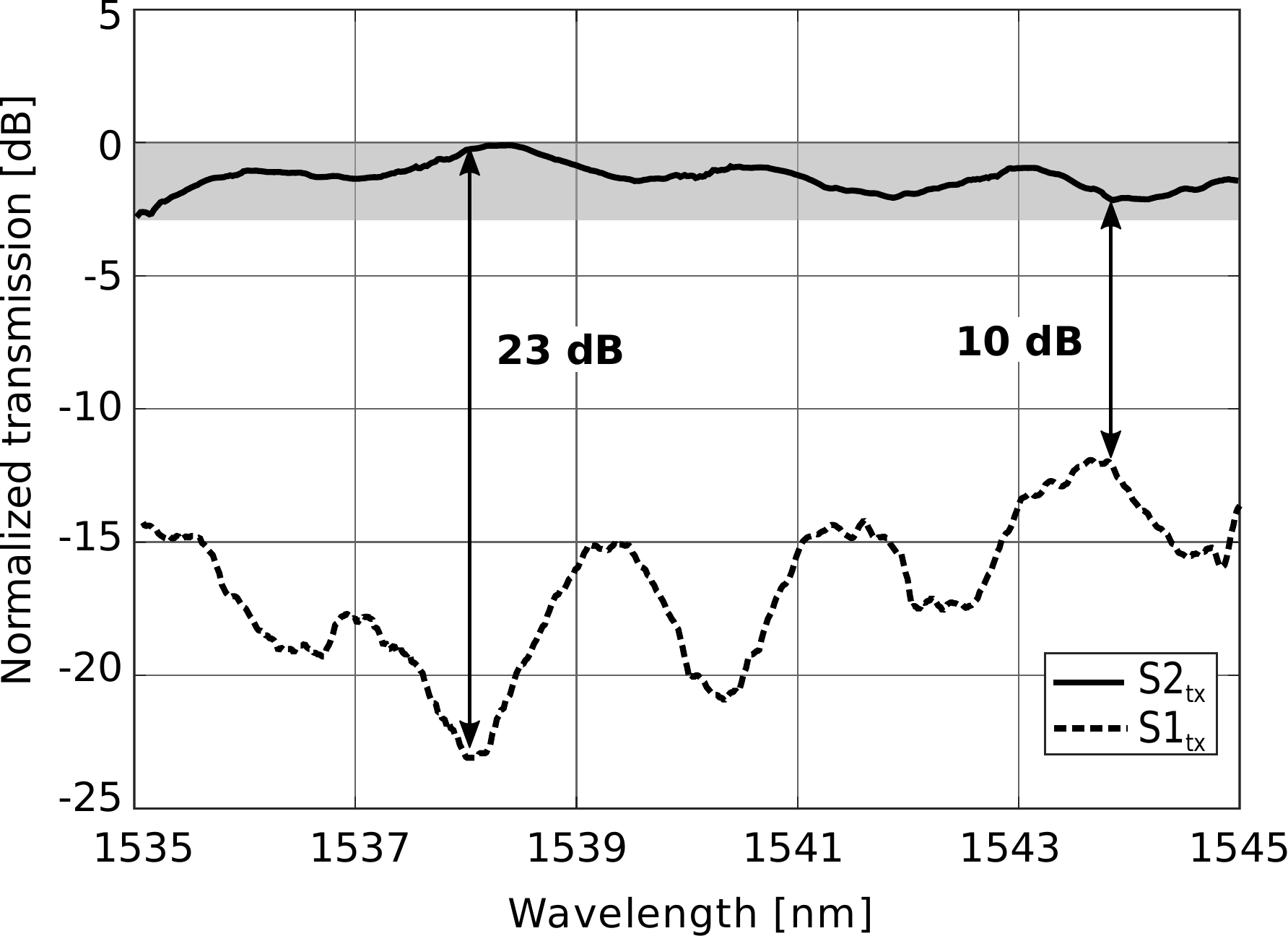}
\caption{Frequency response at S2$_{\mathrm{rx}}$ for both input S1$_{\mathrm{tx}}$ and S2$_{\mathrm{tx}}$. Mode MUX and DEMUX are configured in order to demultiplex S2$_{\mathrm{tx}}$ at S2$_{\mathrm{rx}}$. The channel cross-talk is lower than -10 dB on the entire 10-nm bandwidth. Channel attenuation fluctuates less than 2.5 dB (shaded area).}
\label{fig: wavelength sweep}
\end{figure}
\end{center}

\begin{figure*}[ht]
\centering
\includegraphics[keepaspectratio=true, width=2\columnwidth]{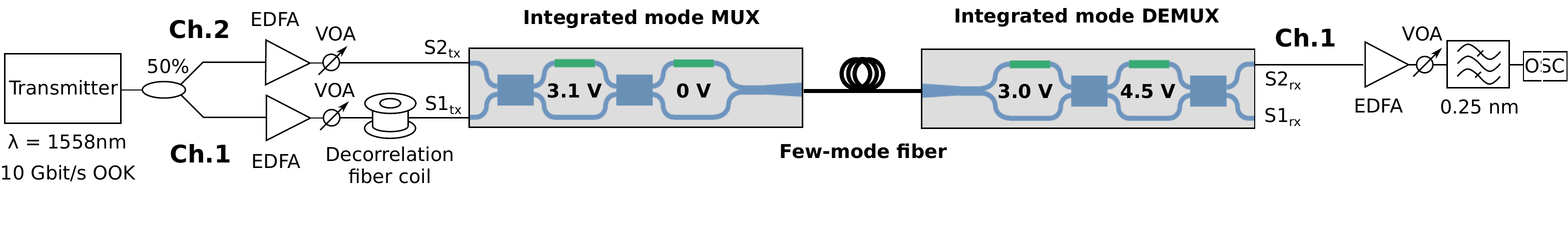}
\caption{Experimental setup for SDM transmission. Two 10Gbit/s OOK NRZ channels are multiplexed on modes LP$_{01}$ and LP$_{11a}$ of the few-mode fiber by means of the integrated mode MUX. Integrated mode DEMUX is used at the receiver.}
\label{fig: sdm setup}
\end{figure*}

Figures \ref{fig:switch maps}(b-e) show the measured transmission from inputs S1$_{\mathrm{tx}}$ and S2$_{\mathrm{tx}}$ to outputs S1$_{\mathrm{rx}}$ and S2$_{\mathrm{rx}}$. Colour scale represents the normalized power received at the output photodiodes. In all considered cases total power loss from MUX inputs to DEMUX outputs is about 30 dB.  %la normal fatta con -30.14 che e max dei due casi
As can be seen, inputs S1$_{\mathrm{tx}}$ and S2$_{\mathrm{tx}}$ can be demultiplexed to either S1$_{\mathrm{rx}}$ or S2$_{\mathrm{rx}}$ exploiting several DEMUX configurations. Considering the case V$_{A,RX}$ = 3.0 V and V$_{P,RX}$ = 4.5 V, channel injected at S1$_{\mathrm{tx}}$ is demultiplexed at S2$_{\mathrm{rx}}$ with an extinction ratio E$_{R,12}$ = 16.3 dB; in this configuration channel at S2$_{\mathrm{tx}}$ is demultiplexed at S1$_{\mathrm{rx}}$ with E$_{R,21}$ = 13.5 dB. A channel cross-talk of -15.4 dB and -14.5 dB is achieved at S1$_{\mathrm{rx}}$ and S2$_{\mathrm{rx}}$, respectively. A differential attenuation of less than 1 dB is achieved between the two channels. The demultiplexing of the input channels can be reversed exploiting DEMUX thermo-optic heaters. Keeping for example the phase controller fixed at V$_{P,RX}$ = 4.5 V while changing the controller of the tunable coupler at voltage V$_{A,RX}$ = 5.2 V, channel at S1$_{\mathrm{tx}}$ is routed to S1$_{\mathrm{rx}}$ with extinction ratio E$_{R,11}$ = 16.3 dB while S2$_{\mathrm{tx}}$ is demultiplexed at S2$_{\mathrm{rx}}$ with E$_{R,22}$ = 14.5 dB. The measured channel cross-talk is -15.2 dB and -15.6 dB at S1$_{\mathrm{rx}}$ and S2$_{\mathrm{rx}}$, respectively. Also in this case differential attenuation between the two channel is smaller than 1 dB. The same demultiplexing can be obtained  also configuring the DEMUX with V$_{A,RX}$ = 3.0 V and V$_{P,RX}$ = 0 (not shown in the figures). Table \ref{tab:table 1} summarises the performance of the integrated (DE)MUX circuits for two demultiplexing configurations.  Demultiplexing reconfiguration can be obtained also exploiting the controllers on the MUX circuit instead of changing the DEMUX settings.

%lo spettro sarebbe 3.1 0 pero' per consistenza con la tabella lo associo a 5.2 4.5
In order to verify the operational bandwidth of the system, the frequency domain response of the SDM transmission was measured at output S2$_{\mathrm{rx}}$. Figure \ref{fig: wavelength sweep} shows the normalized transmission for both inputs when S1$_{\mathrm{tx}}$ and S2$_{\mathrm{tx}}$ are demultiplexed at S1$_{\mathrm{rx}}$ and S2$_{\mathrm{rx}}$, respectively (as in the second column of Table \ref{tab:table 1}). On the measured bandwidth from 1535 nm to 1545 nm the channel attenuation for S2$_{\mathrm{tx}}$ is relatively flat with fluctuations smaller than 2.5 dB (shaded area). Cross-talk (S1$_{\mathrm{tx}}$ transmission) is lower than -10 dB on the entire bandwidth. A cross-talk of less than -20 dB is achieved on a bandwidth of about 100 GHz with central frequency 1538 nm, where it is as low as -23 dB.

%------------------------------------------------
%%%DA QUI%%%
\section{SDM transmission results}
\label{sec: SDM}

Figure \ref{fig: sdm setup} shows the experimental setup used for SDM trasmission exploiting the integrated mode MUX and DEMUX circuits. Two 10 Gbit/s OOK NRZ channels were generated by splitting data stream from a 10G Cisco SFP+ module (carrier wavelength 1558 nm) and decorrelated by a 10-km-long fiber coil. EDFAs and variable optical attenuators were used to ensure the same power level of 0 dBm on both channels at the chip input. Channels 1 and 2 were injected at ports S1$_{\mathrm{tx}}$ and S2$_{\mathrm{tx}}$ of the integrated MUX by means of a fiber array equipped with two standard SMFs without a polarization controller. The few-mode fiber was aligned to MUX and DEMUX chips as described in section \ref{sec: realization}. As in the previous section, mode MUX was set in order to map channel 1 to mode LP$_{11a}$ of the fiber and channel 2 to mode LP$_{01}$ (V$_{\mathrm{A,tx}}$ = 3.1V, V$_{\mathrm{P,tx}}$ = 0V). At the receiver, mode LP$_{11a}$ was demultiplexed at S2$_{\mathrm{rx}}$ and LP$_{01}$ at S1$_{\mathrm{rx}}$ (V$_{\mathrm{A,rx}}$ = 3.0V, V$_{\mathrm{P,rx}}$ = 4.5V). This system configuration is the same of the first column of Table \ref{tab:table 1} and allows to route channel 1 from input S1$_{\mathrm{tx}}$ to output S2$_{\mathrm{rx}}$ and channel 2 from S2$_{\mathrm{tx}}$ to S1$_{\mathrm{rx}}$. At the output of the integrated mode DEMUX, channel 1 was sent into an optical oscilloscope after amplification and filtering (filter bandwidth 0.25 nm). A chip without photodiodes was used for the demultiplexing.

\begin{center}
\begin{figure}[ht]
\centering
\includegraphics[scale=0.6]{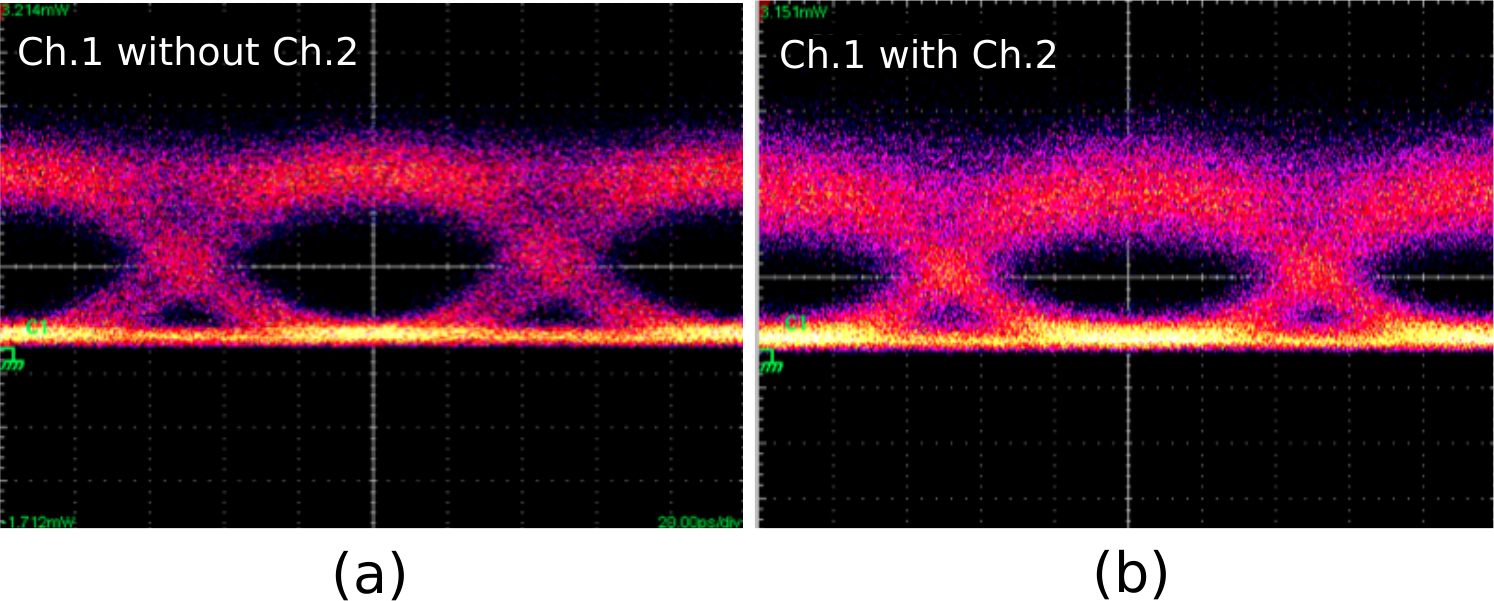}
\caption{Eye diagrams of channel 1 when demultiplexed at output port S2$_{\mathrm{rx}}$ of the mode DEMUX (a) without channel 2 and (b) with simultaneous transmission of channel 2.}
\label{fig: eye diagrams}
\end{figure}
\end{center}

The eye diagram of channel 1 extracted at the output port S2$_{\mathrm{rx}}$ of the mode DEMUX is shown in figure \ref{fig: eye diagrams} when channel 2 is switched off (a) and when channel 2 is switched on (b). As can be seen, a good eye-opening is preserved with only a small signal degradation on the eye diagram (b) when compared to single channel transmission. The measured eye Q-factor reduces from 5.7 (single channel transmission) to 3.8 (double channel transmission). Similar results were obtained switching channel mapping, that is coupling channel 1 to mode LP$_{01}$ of the FMF and the interfering channel 2 to mode LP$_{11a}$.

\section{Conclusion}
An Indium-Phosphide-based photonic integrated circuit for two-mode multiplexing and demultiplexing was presented and experimentally demonstrated. Integrated mode MUX and DEMUX were simultaneously exploited for the transmission of two 10 Gbit/s channels at the same wavelength and polarization over modes LP$_{01}$ and LP$_{11a}$ of a few-mode fiber. Integrated devices present a good coupling efficiency with fiber modes with differential losses smaller than 1 dB. A mode excitation cross-talk of -20 dB and a channel cross-talk suppression after fiber propagation and demultiplexing of 15 dB were measured. An operational bandwidth of the transmission system of at least 10 nm was demonstrated. Both mode MUX and DEMUX are fully reconfigurable and allow a dynamic switch of channel routing in the transmission system. 

% or


%\appendices


\section*{Acknowledgment}

The authors gratefully acknowledge OFS (Denmark) for providing the few-mode fiber, Chigo Okonkwo (COBRA Institute, Eindhoven University of Technology) and Francesco Morichetti (Politecnico di Milano) for precious suggestions and hints and Nicola Peserico (Politecnico di Milano) for help in the experiments. The authors thank Franscisco M. Soares, Moritz Baier and Norbert Grote (Fraunhofer Heinrich Hertz Institut) for the fabrication of the devices.

\ifCLASSOPTIONcaptionsoff
  \newpage
\fi



\bibliographystyle{IEEEtran}
\bibliography{Article_MultiMode}
%\begin{thebibliography}{1}
%\end{thebibliography}


%\newpage


%\vfill


\end{document}